\documentclass[journal=jacsat,manuscript=article]{achemso}

\usepackage[version=3]{mhchem} 
\usepackage{xcolor}
\usepackage{booktabs}
\usepackage[numbers]{natbib}    
\usepackage{graphicx}          
\usepackage{float}
\usepackage{subcaption}
\usepackage[normalem]{ulem}
\usepackage{lipsum}  
\usepackage{stackengine}

\author{Archisman Sinha}
\affiliation{School of Chemical Sciences, Indian Association for the Cultivation of Science, Kolkata-700032, India}

\author{Pintu Mandal}
\email{pintuphys@gmail.com}
\affiliation{Department of Physics, St. Paul's Cathedral Mission College, Kolkata-700009, India}

\author{Nabanita Deb}
\email{nabanita.deb@iacs.res.in}
\affiliation{School of Chemical Sciences, Indian Association for the Cultivation of Science, Kolkata-700032, India}

\title[An \textsf{achemso} demo]
 {Anomalous Peak Formation in the Second Stability Zone of Quadrupole Mass Filters: Role of Non-linear Resonances Induced by Single Rod Defects}

\abbreviations{}

\begin{document}

\newpage
\begin{abstract}
Operating a quadrupole mass filter (QMF) within its second stability zone offers superior mass resolving power but introduces extreme sensitivity to structural imperfections. While symmetric misalignments are well-documented, this work combines analytical modeling with SIMION trajectory simulations to investigate the unaddressed electrodynamic impact of a completely localized, asymmetric defect on a single lone rod. Unlike diagonally symmetric perturbations, which reduce the four-fold rotational symmetry to two-fold symmetry while maintaining a smooth stability landscape, a single-rod defect breaks the remaining symmetry, giving rise to pronounced transmission ridges within the second stability zone. Spatial multipole expansion reveals that this structural breakdown injects odd-parity harmonics—dominated by the hexapole ($A_3$) field—which couple the orthogonal transverse equations of motion. By tracking secular indices through explicit zone-II Floquet projection mappings, we demonstrate that these asymmetric fields drive destructive, higher-order nonlinear secular resonances. Ions traversing these precise parametric coordinates undergo rapid amplitude growth and collide with adjacent electrodes, establishing critical geometric tolerance frameworks and operating conditions for high-performance mass spectrometry.  
\end{abstract}


\begin{tocentry}
    \centering
    \includegraphics[width=8.25cm, height=4.45cm]{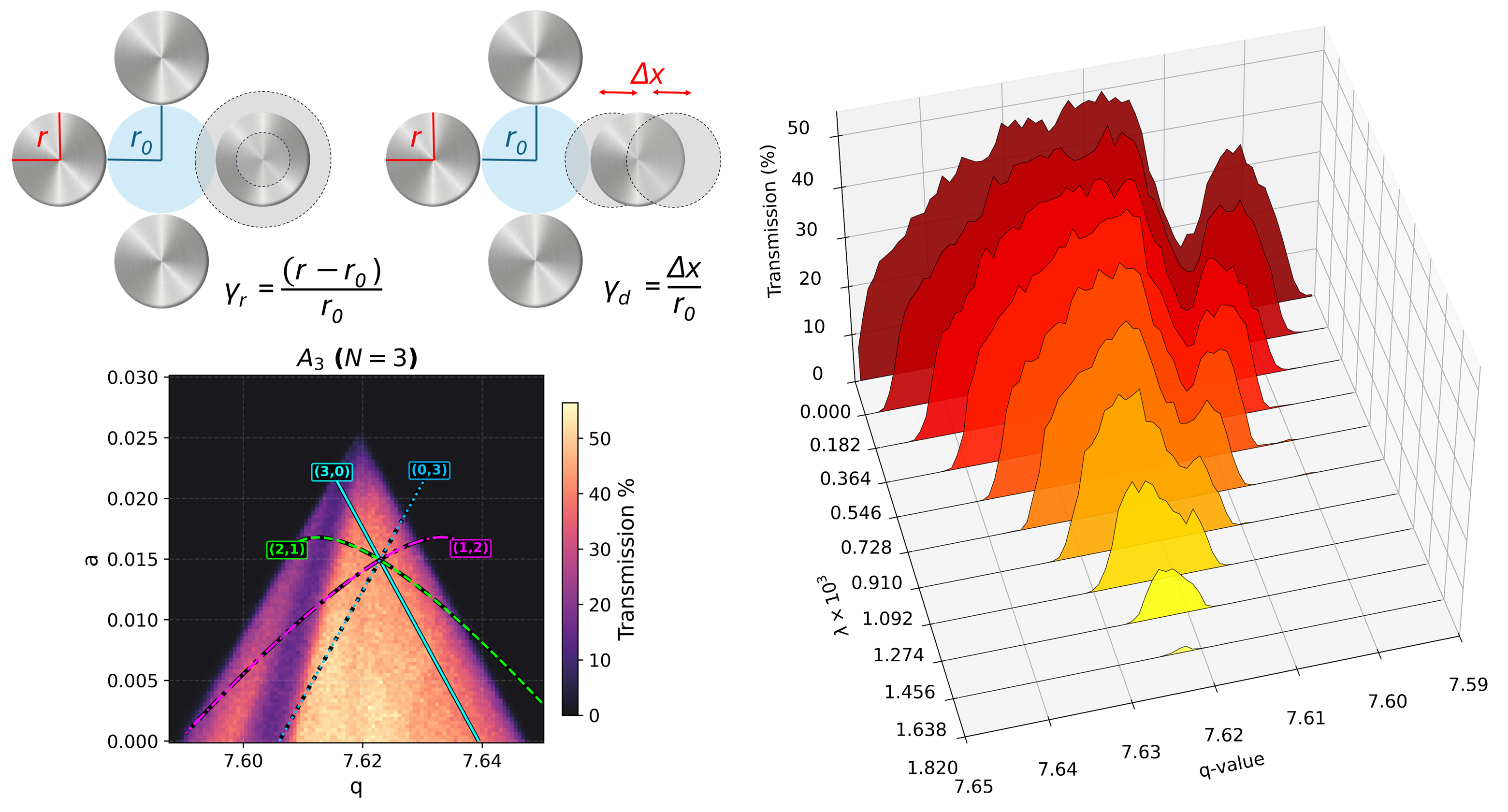}
\end{tocentry}

\newpage
\section{Introduction}

Since its foundational introduction by Wolfgang Paul and co-workers,\cite{Paul1953, Paul1955, paul1990electromagnetic} the conventional linear radio-frequency (RF) quadrupole mass filter (QMF) has become an indispensable cornerstone of atomic, molecular, and industrial chemical physics.
In an ideal, infinitely long quadrupole mass filter, four parallel electrodes with perfect hyperbolic cross-sections generate a pure two-dimensional quadrupole potential~\cite{Paul1953}. Under these ideal conditions, ion motion through the field is entirely decoupled along the orthogonal transverse axes ($x$ and $y$) and can be elegantly modeled via classical linear Mathieu differential equations~\cite{McLachlan1951}.
Stable coordinates within the resulting two-dimensional $(a,q)$ parameter space delineate operating regions where an ion maintains bounded amplitudes over infinite RF periods.
This allows the ion to successfully traverse the analyzer to a downstream detector without colliding with the electrode boundaries.

Historically, commercial instruments operate near the primary apex of the first stability zone ($a \approx 0.23, q \approx 0.706$)\cite{Miller1986} to ensure uniform transmission efficiency and straightforward mass scanning lines at lower operating voltages. Nevertheless, growing analytical demands across complex biological macromolecular structures~\cite{groetsema2025accessing}, space exploration missions~\cite{houk1980inductively}, and ultra-high-resolution isotope isolation~\cite{ying1996high, chen2000high, hu2024simulation} have driven a modern renaissance in exploring higher-order Mathieu stability fields~\cite{chakravorty2024experimental, reinsfelder1976phd, reinsfelder1981theory, hogan2008performance}.
In particular, operation within the narrow boundaries of the second stability zone\cite{Dawson1974, Dawson1976} ($7.51 < q < 7.58$) yields unparalleled resolving power over short RF exposures, alongside higher acceptable ion beam kinetic energies owing to an expanded pseudopotential well depth.
Crucially, the inherent architectural narrowness of this second stability region makes it exceptionally sensitive to subtle manufacturing variances and structural non-idealities.

To achieve structural fabrication feasibility and ease of mechanical assembly, standard commercial and experimental instruments typically substitute ideal hyperbolic profiles with parallel cylindrical rods.\cite{Dayton1954, Denison1971, cheng2024planar} Although this geometry preserves the four-fold rotational symmetry of the electrode arrangement, the desired quadrupole component ($A_2$) is accompanied by intrinsic higher-order even multipole components ($A_6$, $A_{10}$, $\ldots$) permitted by the symmetry condition $N = 2(2m+1)$.\cite{Douglas2014}
Optimizing the rod-to-field radius ratio ($\eta = r/r_0 \approx 1.13$) minimizes the dominant intrinsic distortion, namely the dodecapole component ($A_6$), while the remaining higher-order even multipoles persist.\cite{Gibson2000, Konenkov2002, douglas2002influence} Breaking the four-fold rotational symmetry of the cylindrical-rod quadrupole geometry fundamentally alters the multipole structure of the field. A reduction to two-fold symmetry, as produced by symmetric diagonal perturbations, introduces additional even-order multipoles, most notably the octupole ($A_4$), while preserving the absence of odd harmonics~\cite{Dutta2026,Jana2025RadialAI}. In contrast, complete symmetry breaking arising from a localized single-rod defect permits the emergence of odd-order multipoles, dominated by the hexapole ($A_3$), in addition to the inherent even-order components.\cite{Jana2025Effect, Taraphdar2026, douglas2003octopole, konenkov2010hexapole} Although the electrodynamic consequences of symmetry breaking have been extensively investigated in the first stability zone~\cite{Jana2025Effect, Jana2025RadialAI, Taraphdar2026}, and symmetric diagonal perturbations have recently been examined in the second stability zone,\cite{Dutta2026} the influence of localized single-rod defects on second-zone ion dynamics remains unexplored.

In this paper, we deploy a unified combination of analytical electrodynamic models and numerical ion flight profiling via SIMION to explore this configuration.
Through precise multipole fitting expansion models, we track the evolution of field distortions arising from a single-rod perturbation.
Most notably, we map out a severe transmission valley structure—a localized dip or ridge running directly across the interior of the second stability map.
Using fundamental secular index assessments ($\beta_x, \beta_y$), we provide a definitive physical explanation for this transmission drop, showing it corresponds to higher-order nonlinear resonances driven by the newly injected octupolar and hexapolar structural fields.\cite{Du1999}
Finally, we illustrate how these transmission dips are related to the non-linear resonance (NLR) lines, providing a robust engineering framework to bypass these performance-limiting field traps.


\section{Theory and Simulation Methods}

The electrodynamic confinement within a linear QMF is achieved by applying DC and RF voltages of opposite polarity to the two diagonal electrode pairs, thereby generating the ideal two-dimensional quadrupole potential distribution,
\begin{equation}
    \Phi(x, y, t) = \frac{x^2 - y^2}{r_0^2} \left(U - V \cos\Omega t \right)
\end{equation}
where $U$ is the applied DC voltage, $V$ is the amplitude of the RF voltage, and $\Omega = 2\pi f$ is the driving angular frequency.
Under a dimensionless time scale transformation $\xi = \Omega t / 2$, the transverse trajectories of an ion of mass $m$ and charge $e$ reduce to the uncoupled canonical Mathieu equations:\cite{McLachlan1951}
\begin{equation}
    \frac{d^2 u}{d\xi^2} + \left[ a_u - 2q_u \cos(2\xi) \right] u = 0, \quad (u = x, y)
\end{equation}
where the dimensionless Mathieu operating coordinates scale as $a_x = -a_y = \frac{8eU}{m\Omega^2 r_0^2}$ and $q_x = -q_y = \frac{4eV}{m\Omega^2 r_0^2}$.
According to Floquet's theorem, stable solutions are formulated as non-vanishing periodic modulations:\cite{McLachlan1951, douglas2020phasemodulation}
\begin{equation}
    u(\xi) = \alpha'_u e^{i\beta_u \xi} \sum_{n} C_{2n, u} e^{2in\xi} + \alpha''_u e^{-i\beta_u \xi} \sum_{n} C_{2n, u} e^{-2in\xi}
\end{equation}
where the fundamental bounded states are determined by a real fractional secular index $\beta_u \in (0,1)$ in the first stability zone.\cite{Dawson1976}

\subsection{Modified Stability Zone}
Localized mechanical aberrations, such as a radius variation of a pair of rods, disrupt the four-fold rotational mirror symmetry and perturb the ideal quadrupole potential~\cite{Jana2025RadialAI}. For a dual-rod asymmetry characterized by the normalized scaling factor $\gamma$, the resulting geometric distortion modifies the field structural constants, leading to effective operating parameters $(a_{\text{eff}}, q_{\text{eff}})$ scaled by a geometric factor $p$:\cite{Jana2025RadialAI}
\begin{equation}
    p = \frac{1 + (1 + \gamma)^2}{2}
    \label{eq: p factor}
\end{equation}

\begin{figure}[h] 
  \centering
  \includegraphics[width=1.0\textwidth]{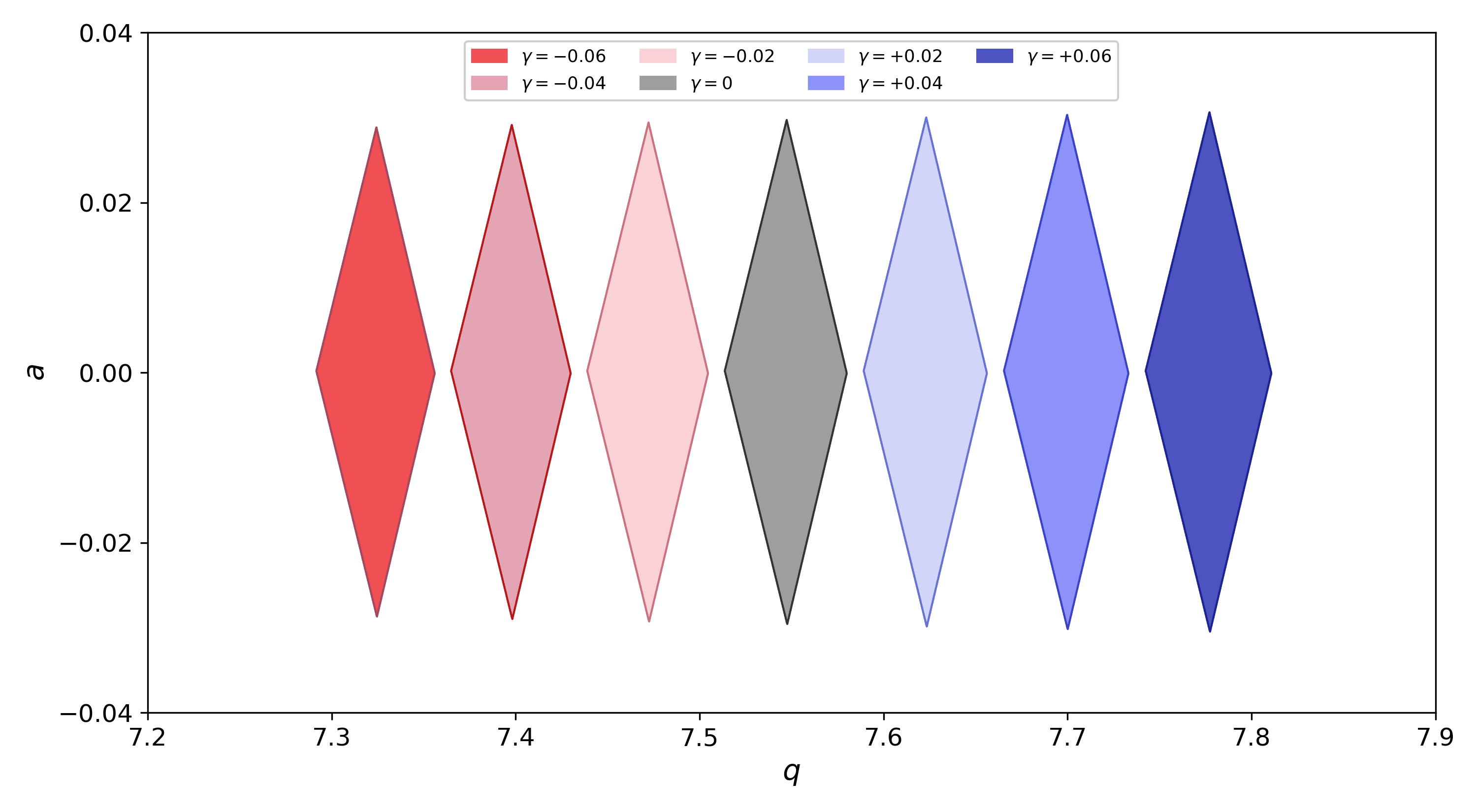}
  \caption{Analytical stability diagram scaled using the $p$-factor for a single rod perturbation where $\gamma$ is the asymmetry parameter.}
  \label{fig:analytical_stability_diagram}
\end{figure}

For a single-rod asymmetry, the effective geometric distortion is reduced by a factor of `two', relative to the dual-rod configuration. Consequently, the quadrupole potential can be equivalently described by replacing the perturbation parameter ($\gamma$) with ($\gamma/2$) in eq.~\ref{eq: p factor}. In Figure~\ref{fig:analytical_stability_diagram} the analytical stability diagram has been derived for the single rod asymmetry cases by considering this approximation. The stability grids were mapped out via a high-performance, parallelized 4th-order Runge-Kutta (RK4) scheme~\cite{Jana2025RadialAI}. Trajectories were initialized with a displacement of $x_0, y_0 = 0$ and initial velocities of $v_{x0}, v_{y0} = 1$, and integrated over a long dimensionless time interval ($\xi_{\text{max}} = 150$) with a discrete step size of $d\xi = 0.05$. A coordinate pair $(q, a)$ was classified as unstable if the displacement or velocity along either orthogonal axis exceeded a predefined bound ($x, y > 200$) at any point during the execution. The apex of the stability zone shifts in the $q$-parameter space with increasing asymmetry parameter, in good agreement with the simulated transmission profiles and stability diagrams obtained from SIMION, as shown in the subsequent section. The only feature this diagram cannot capture is the realistic transmission loss due to the non-linear field components introduced by the modified geometry.

\subsection{Secular Indices}
The infiltration of high-order multipole spatial harmonics couples the transverse coordinates and introduces nonlinear forcing fields.
To isolate the background secular frequency indices $\beta_x$ and $\beta_y$ across a dense parameter meshgrid $(a, q)$ within this asymmetric field geometry, Floquet method must be applied using these effective coordinates:\cite{McLachlan1951, Konenkov2002}
\begin{equation}
    a_{\text{eff}} = \frac{a}{p}, \quad q_{\text{eff}} = \frac{q}{p}
    \label{eq:effective_parameters}
\end{equation}

To avoid fractional ambiguity and preserve continuous root tracking across these higher stability boundaries, the real-valued secular frequency components must be converted via the explicit zone-II projection mappings:\cite{Du1999, Konenkov2002} For determining the secular frequencies in the time domain, the equations of motion for each transverse axis $u \in \{x, y\}$ are integrated across exactly one period of the radiofrequency drive ($\Delta \xi = \pi$).
This mapping defines the $2 \times 2$ Floquet transfer matrix (or matrix propagator) $\mathbf{M}_u$, which projects the ion's initial state vector onto its configuration at the end of the RF cycle:
\begin{equation}
    \begin{pmatrix} u(\pi) \\ u'(\pi) \end{pmatrix} = \mathbf{M}_u \begin{pmatrix} u(0) \\ u'(0) \end{pmatrix}
\end{equation}
where the individual columns of $\mathbf{M}_u$ are constructed by numerically propagating two linearly independent sets of initial conditions, representing fundamental solutions $u_1(\xi)$ and $u_2(\xi)$, through the state equations:
\begin{equation}
    \mathbf{M}_u = \begin{pmatrix} u_1(\pi) & u_2(\pi) \\ u_1'(\pi) & u_2'(\pi) \end{pmatrix}, \quad \text{with} \quad \begin{pmatrix} u_1(0) \\ u_1'(0) \end{pmatrix} = \begin{pmatrix} 1 \\ 0 
\end{pmatrix}, \ \begin{pmatrix} u_2(0) \\ u_2'(0) \end{pmatrix} = \begin{pmatrix} 0 \\ 1 \end{pmatrix}
\end{equation}

By invoking Liouville's theorem for conservative systems ($\det \mathbf{M}_u = 1$), the eigenvalues of the propagator take the form $e^{\pm i \pi \beta_u}$.
The trace of the transfer matrix ($\text{Tr}(\mathbf{M}_u) = u_1(\pi) + u_2'(\pi)$) directly isolates the fractional phase shift, yielding the principal baseline exponent:
\begin{equation}
    \beta_{u,\text{principal}} = \frac{1}{\pi} \arccos \left[ \frac{\text{Tr}(\mathbf{M}_u)}{2} \right]
\end{equation}

However, when tracking analytical lines into the highly selective second stability zone ($7.51 < q < 7.58$), the highly oscillatory trajectories undergo phase shifts that cross the fundamental integer boundary.
To map the principal components back into the physical branches of the second stability zone without fractional ambiguity, the asymmetry of the operating coordinates ($a_{\text{eff}, x} = a_{\text{eff}}$ and $a_{\text{eff}, y} = -a_{\text{eff}}$) requires explicit zone-II projection mappings:\cite{Du1999, Konenkov2002}
\begin{equation}
    \beta_x = 2 - \beta_{x,\text{principal}}(a_{\text{eff}}, q_{\text{eff}})
\end{equation}
\begin{equation}
    \beta_y = 2 - \beta_{y,\text{principal}}(-a_{\text{eff}}, q_{\text{eff}})
\end{equation}

By computing this numerical propagator across the parameter expanse of the second stability zone, this theoretical framework isolates the operational coordinates where the ions meet the NLR lines~\cite{Du1999},
thereby giving rise to the characteristic transmission ridges and dips observed in transmission profiles.

\subsection{Multipole Field Analysis}
To accurately characterize the potential profiles within the non-ideal QMF geometry, the spatial potential distribution was decomposed into a series of orthogonal 2D multipole fields.\cite{Douglas2014, Konenkov2002}
This decomposition allows for the systematic isolation of the dominant quadrupolar component alongside higher-order parasitic multipole terms arising from geometric asymmetries or non-ideal electrode configurations. The electrostatic potential $\Phi(x, y)$ in a charge-free region can be modeled using a complex multipole expansion. Specifically, a static potential difference was established by biasing the $x$-pair rods to $+1\text{ V}$ and the $y$-pair rods to $-1\text{ V DC}$. The resulting electrostatic potentials within the central region of the QMF were then sampled across a Cartesian $(x, y)$ grid with a spatial resolution of $0.1\text{ units}$, bounded circularly by the inscribed field radius $r_0$.
By mapping the Cartesian coordinates $(x, y)$ onto the complex plane via $z = x + iy$, the potential field is expanded up to a maximum order of $N_{\max}$.
The normalized complex spatial variable is defined relative to the inscribed field radius $r_0$ ($r_0 = 5.0$~units) as:
\begin{equation}
\Phi(x, y) = \sum_{N=0}^{N_{\max}} \left[ a_N \cdot \text{Re}\left(\left(\frac{z}{r_0}\right)^N\right) -   b_N \cdot \text{Im}\left(\left(\frac{z}{r_0}\right)^N\right) \right]
\end{equation}
where $a_N$ represents the normal (axis-aligned) multipole coefficients, $b_N$ represents the skew (rotated) multipole coefficients~\cite{douglas2009linear,douglas1999spatial,konenkov2010spatial}, and $N_{\max} = 14$ defines the truncation limit for high-order field considerations.
Due to the physical dominance of the quadrupolar field in a functional instrument, the initial guess vector ($p_0$) was heavily weighted toward the principal normal quadrupole coefficient by setting $a_2 = 1.0$, while initializing all other coefficients to $0.0$ to ensure rapid convergence and avoid local minima.
Following the extraction of individual normal ($a_N$) and skew ($b_N$) parameters for every order, a co-ordinate independent signed magnitude representation ($A_N$) was reconstructed:
\begin{equation}
A_N = \text{sgn} \cdot \sqrt{a_N^2 + b_N^2}
\end{equation}
The sign convention ($\text{sgn}$) for the ultimate coefficient was determined by assessing the dominant spatial orientation component:
\begin{equation}
\text{sgn} = \begin{cases} 
\text{sign}(a_N), & \text{if } |a_N|
\ge |b_N| \text{ and } |a_N| > 10^{-9} \\
1.0, & \text{if } |a_N| \ge |b_N| \text{ and } |a_N|
\le 10^{-9} \\
\text{sign}(b_N), & \text{if } |a_N| < |b_N|
\end{cases}
\end{equation}

This technique isolates the net energy contribution of each multipole order while preserving structural polarity across varying structural asymmetries. 



\subsection{SIMION Setup}

To systematically analyze the electrodynamic consequences of localized geometric non-idealities, a series of virtual high-order mass filter assemblies were constructed using the commercial ion optics simulation software suite SIMION 3D (version 8.2).\cite{Manura2020}
The baseline unperturbed reference design modeled a sinusoidal quadrupole mass filter (QMF) equipped with identical parallel cylindrical rods of nominal radius $r_n$ held at an inscribed field clearance radius of $r_0 = 5.0\text{ mm}$, yielding an optimized structural baseline ratio of $r_n/r_0 \approx 1.13$.
To ensure a fine spatial resolution for the finite difference potential solver within SIMION, the internal geometric coordinates were mapped onto a highly refined mesh grid scaling at an allocation density of $0.1\text{ mm}$ per grid unit, effectively preventing mathematical blurring or boundary discretization artifacts near the circular rod perimeters.

Transverse structural perturbations were then systematically introduced via two independent physical modeling frameworks across an expanse of positive and negative scaling ranges.
In the first simulation series, a localized radial thickness asymmetry was modeled by altering the cross-sectional radius of a single independent rod ($r_1$), making it either larger or smaller than the other three electrodes according to the normalized parameter variation $\gamma_r = (r_1 - r_n)/r_0$.
In the second simulation series, a localized translation imperfection was evaluated by mechanically shifting a single lone rod along its primary axis of focus, generating a discrete inward or outward spatial displacement ($\Delta x$) mapped via $\gamma_d = \Delta x/r_0$.

To establish analytical consistency, benchmark structural configurations against symmetric multipole baselines, and isolate any single-rod coupling effects, a dual-rod thickness control trial was simultaneously mapped out;
this baseline control configuration introduced matching radial thickness variances ($\pm \gamma_r$) onto a diagonally opposite electrode rod pair simultaneously, providing an isolated look at pure even-order field transformations.
The electrical properties of these asymmetric arrays were mapped within the second stability zone boundary matrices ($7.51 < q < 7.58$) by setting up custom sinusoidal voltage waveforms using the SIMION user programming workbench.
Opposing electrode pairs were driven with identical alternating current (RF) voltages operating at a high reference frequency of $f = 0.5\text{ MHz}$, alongside an independently adjusted direct current (DC) biasing offset $U$ to select specific operating points on the $(a,q)$ meshgrid.

Ion transmission characteristics and stability alterations were probed by launching large ensembles consisting of $1000$ to $2000$ singly charged test ions ($m/z = 40\text{ u/C}$) from an entry plane located $1.0\text{ mm}$ upstream of the main quadrupole entrance.
The initial injection conditions assumed a uniform beam configuration characterized by a cross-sectional entry diameter of $r_a = 0.02 r_0$ and an axial forward kinetic energy of $E_k = 2.0\text{ eV}$.
To ensure comprehensive phase spacing coverage, the ion generation timelines were distributed uniformly across 0 to 2 $\mu s$.
Bounded ion trajectories were continuously monitored over an analytical length of $160\text{ mm}$, corresponding to approximately 20 full RF cycles.
A transmitted event was recorded if the ion cleanly passed through the whole QMF length without colliding with the electrode boundaries.
Finally, the transmission intensities were processed into two-dimensional heat maps over the parameter meshgrid to locate the exact parameters where the localized single-rod structural breaking triggers nonlinear secular resonance valleys.


\section{Results and Discussion}

\subsection{Anomalous Transmission Peak}

To evaluate the impact of geometric faults on the mass filter's performance, ion trajectories were simulated utilizing SIMION 3D. The simulations were specifically configured to target the second stability region of the Mathieu stability diagram, localized near a purely theoretical $q$-value of 7.54 for the unperturbed field. For each geometric configuration, an ion ensemble was generated and flown through the QMF. The standard simulations utilized 2000 ions per scan step. The simulations relied on a customized scanning protocol where the dimensionless operating parameters $a$ and $q$ were incrementally stepped while keeping the ration $\lambda=a/2q$ fixed. At each $q$-value, transmitted ions reaching the detector plane were logged. Data extraction scripts paired the recorded $q$-values with the corresponding successful ion counts to derive the absolute transmission percentage.

\begin{figure}[H]
    \centering
    
    \begin{subfigure}{0.85\textwidth}
        \centering
        \stackinset{l}{-0.2cm}{t}{0.2cm}{\textbf{(a)}}{%
            \includegraphics[width=\textwidth]{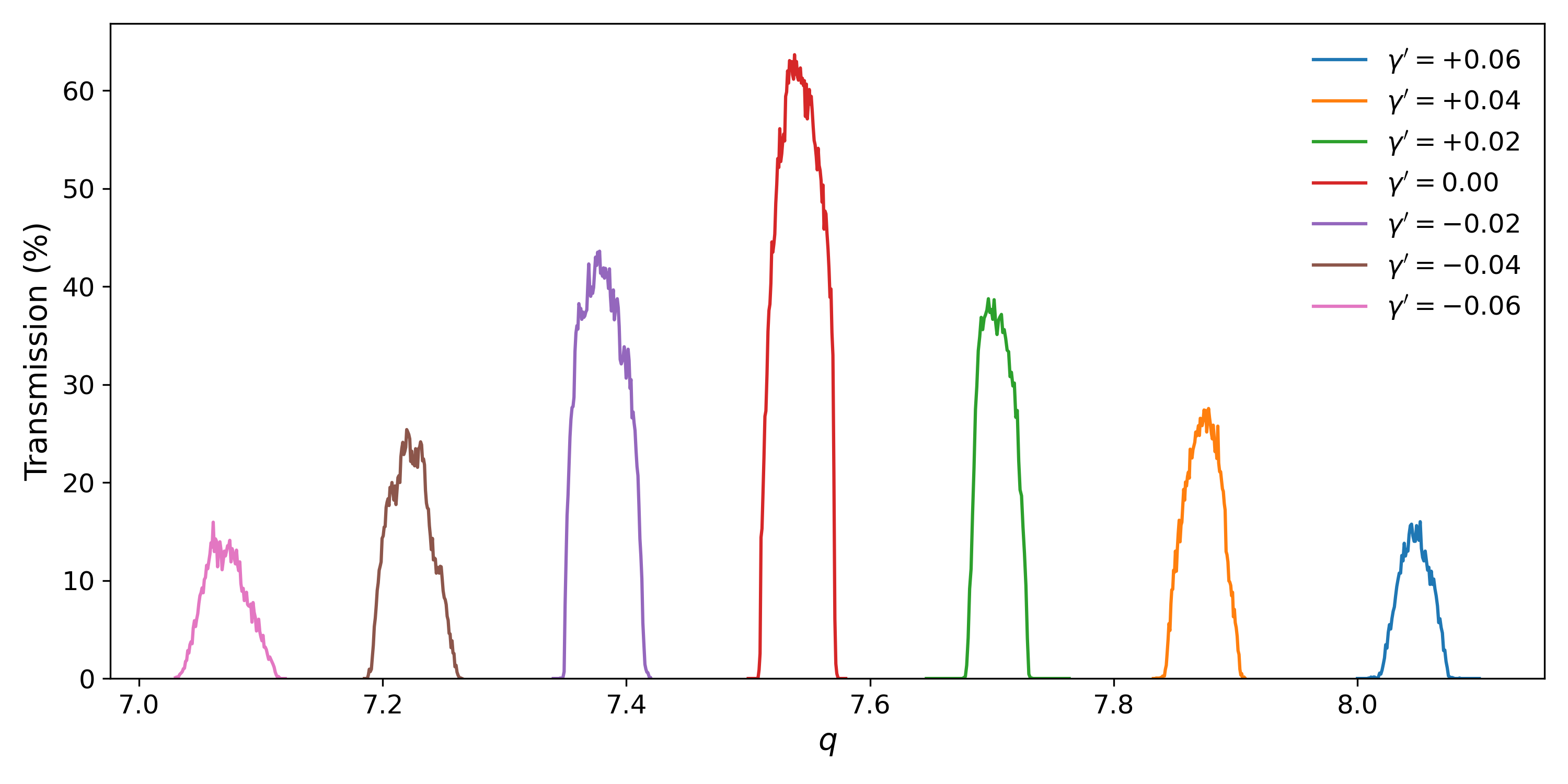}%
        }
        \caption*{} 
        \label{fig:drr_normal}
    \end{subfigure}
    
    \vspace{-0.8cm} 
    
    \begin{subfigure}{0.85\textwidth}
        \centering
        \stackinset{l}{-0.2cm}{t}{0.2cm}{\textbf{(b)}}{%
            \includegraphics[width=\textwidth]{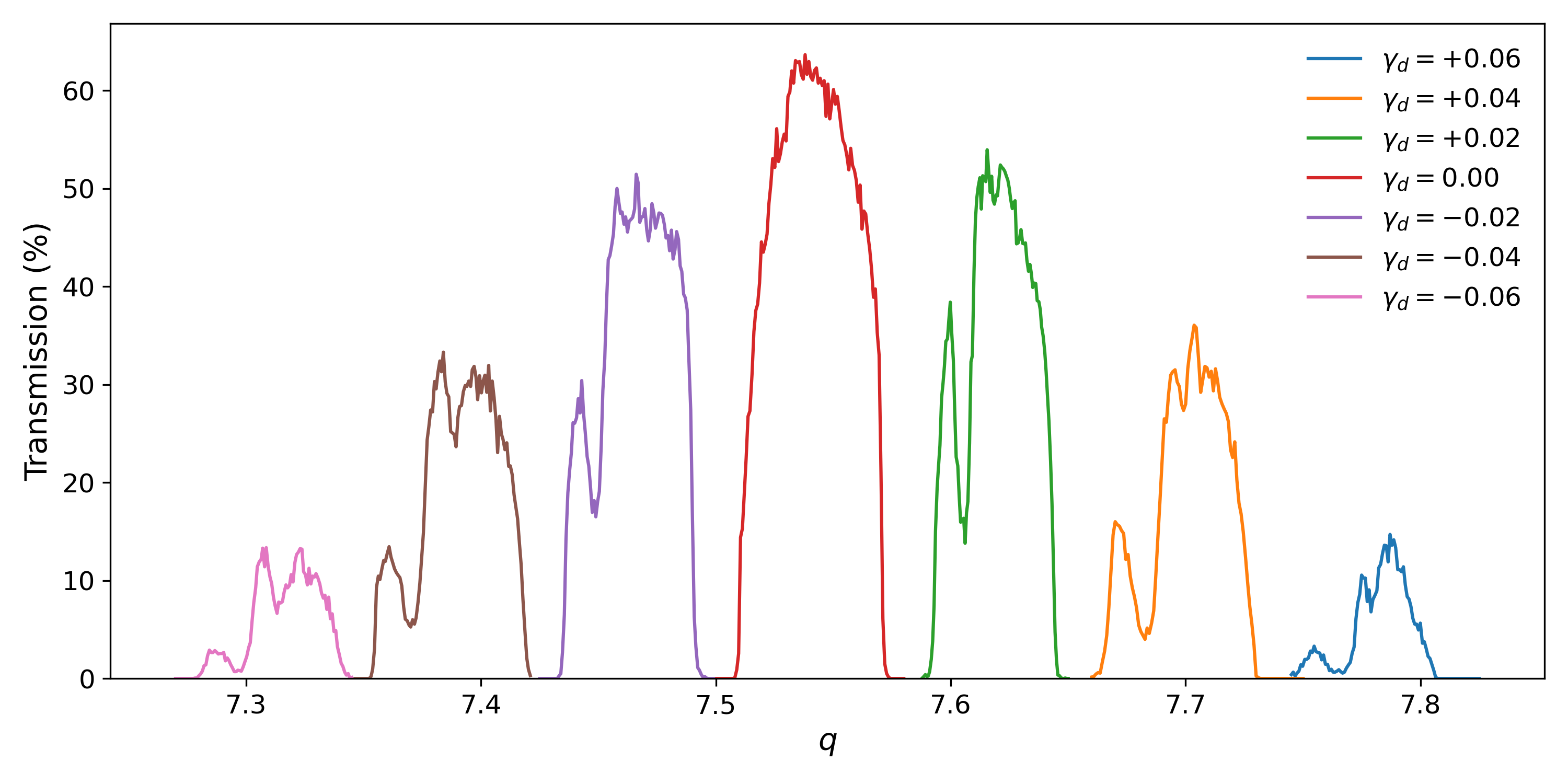}%
        }
        \caption*{} 
        \label{fig:srd_normal}
    \end{subfigure}
    
    \vspace{-0.8cm}
    
    \begin{subfigure}{0.85\textwidth}
        \centering
        \stackinset{l}{-0.2cm}{t}{0.2cm}{\textbf{(c)}}{%
            \includegraphics[width=\textwidth]{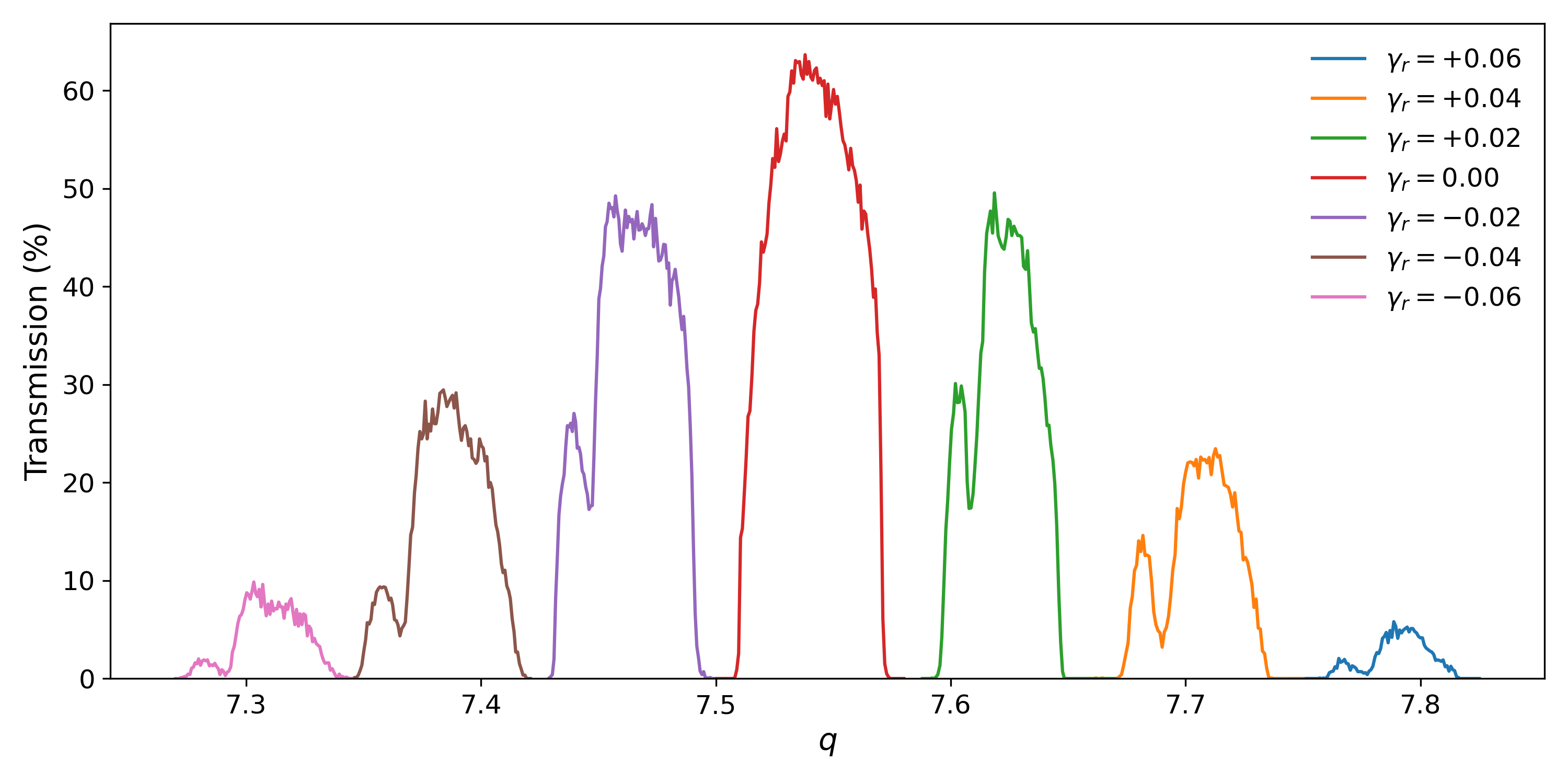}%
        }
        \caption*{} 
        \label{fig:srr_normal}
    \end{subfigure}

    \vspace{-0.8 cm}

    \caption{Transmission efficiency (\%) as a function of $q$-value under X pair rods at (+ DC) and Y pair at (- DC) conditions across different asymmetry configurations ($\gamma$) for (a) dual rod radius asymmetry (DRR); (b) single rod displacement asymmetry (SRD); and (c) single rod radius asymmetry (SRR) cases, using $\lambda=0.000182$ scan line.}
    \label{fig:combined_transmission_normal}
\end{figure}

The unperturbed transmission profile of the QMF with perfect quadrupolar symmetry ($\gamma = 0.00$) yields a characteristic, narrow transmission window centered near $q = 7.54$, achieving a maximum transmission efficiency of approximately 63\%. The introduction of geometrical asymmetries drastically alters both the position and the internal structure of these stability windows. Figures \ref{fig:combined_transmission_normal} (a), \ref{fig:combined_transmission_normal} (b), and \ref{fig:combined_transmission_normal} (c) illustrate the transmission profiles for Dual Rod Radius (DRR) ($\gamma'$), Single Rod Displacement (SRD) ($\gamma_d$), and Single Rod Radius (SRR) ($\gamma_r$) asymmetries, respectively. In each case, the perturbed electrode(s) are biased at the positive DC potential. Across all three fault types, a primary consequence of the asymmetry is a translation of the stability window along the $q$-axis, in agreement with the modified stability diagrams obtained using the fourth-order Runge–Kutta (RK4) method. Positive asymmetry parameters induce a shift toward higher $q$-values, whereas negative parameters shift the window to lower $q$-values, accompanied by a monotonic decrease in overall transmission efficiency.

A stark contrast emerges when comparing the internal structure of the transmission peaks between dual-rod and single-rod faults. In the DRR configuration (Figure \ref{fig:combined_transmission_normal} (a)), the transmission windows maintain a relatively cohesive and smooth profile, without any precursor peak even at strong perturbation ($|\gamma'| = 0.06$). This behavior is consistent with our earlier study on symmetric diagonal rod displacements, which likewise showed no anomalous transmission features despite the reduction of the four-fold symmetry to two-fold symmetry.~\cite{Dutta2026} Conversely, the SRD and SRR configurations (Figures \ref{fig:combined_transmission_normal} (b) and \ref{fig:combined_transmission_normal} (c)) exhibit severe peak degradation, characterized by deep, jagged transmission loss bands dissecting the stability windows. Reversing the polarity of the rods also produces similar kind of transmission profiles.

\newpage

\begin{figure}[H]
\centering
\includegraphics[width=1.0\textwidth]{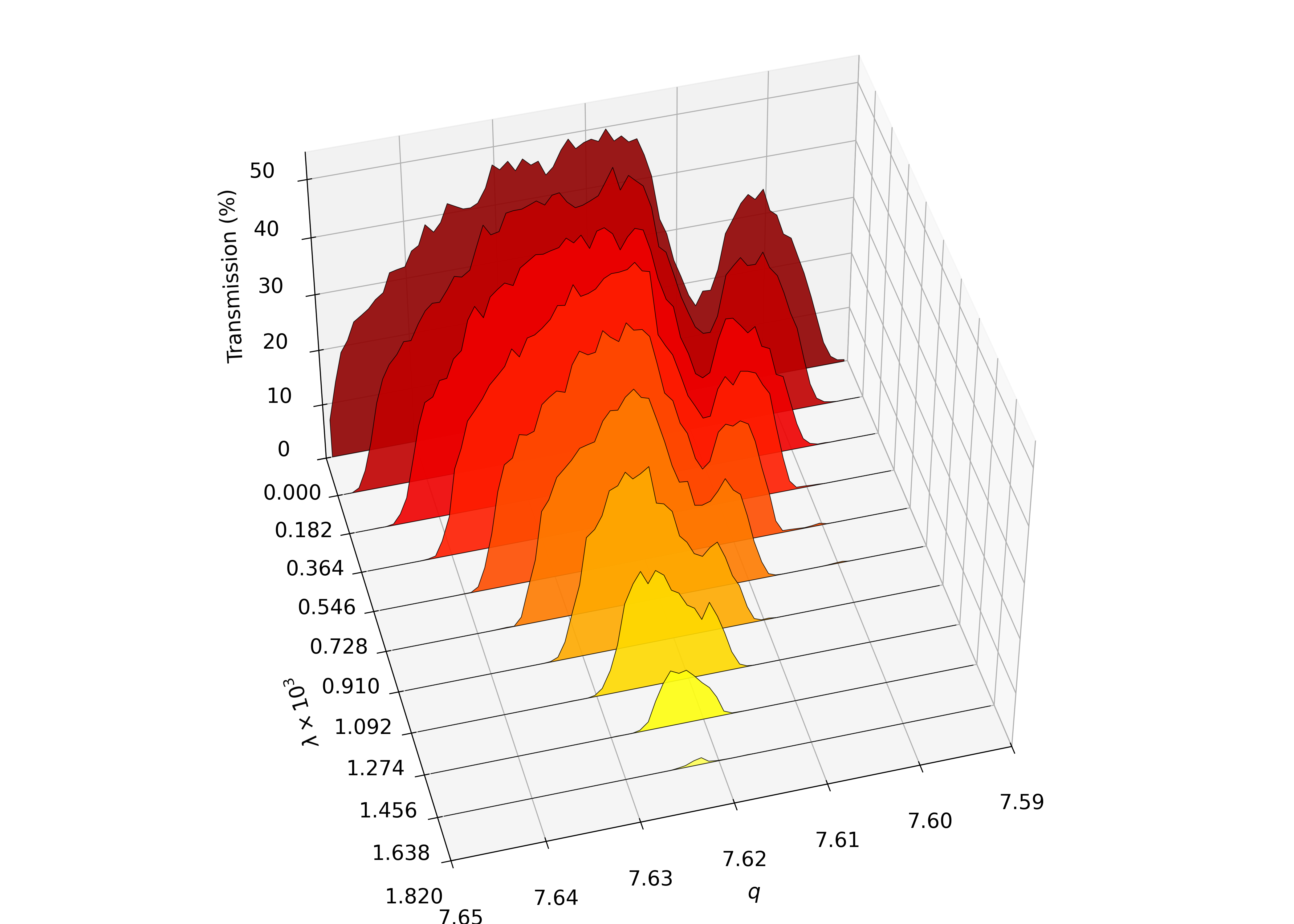}
\caption{SRR transmission scan profiles mapped across varying scan lines ($\lambda$) for $\gamma_r=+0.02$.}
\label{fig:different lambda scan SRR}
\end{figure}

To examine the anomalous peak structure in greater detail, the SRR case at $\gamma_r = +0.02$ was investigated over a range of $\lambda$ scan lines. The corresponding transmission profiles are shown in Figure~\ref{fig:different lambda scan SRR}. As the scan line shifts toward higher values of $\lambda$, the overall transmission decreases. More importantly, the transmission dip gradually weakens and disappears completely at $\lambda = 1.274 \times 10^{-4}$, where the scan line passes close to the apex of the stability region shown in \ref{fig:analytical_stability_diagram}. For all other $\lambda$, however, the dip remains a persistent and reproducible feature unique to the single-rod perturbation, indicating a localized transmission-loss region within the second stability zone rather than an artifact of a particular scan condition. The physical origin of this localized transmission loss is investigated in the following sections. 

\subsection{SIMION Stability Analysis}

\begin{figure}[H]
\centering
\includegraphics[width=1.0\textwidth]{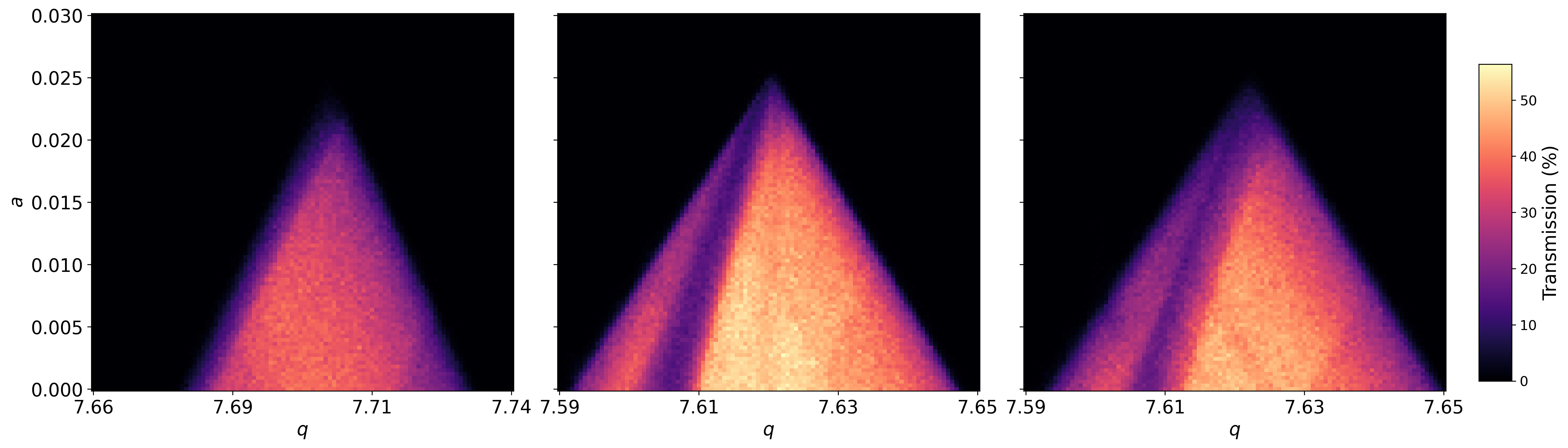}
\caption{Stability diagrams evaluated from SIMION for three different configurations - DRR (left), SRD (middle), SRR (right) (for $\gamma=+0.02$)}
\label{fig:stability diagram comparison}
\end{figure}

To obtain a comprehensive picture of the emergence of the split peaks, SIMION stability diagrams were generated by evaluating the transmission over a $100 \times 100$ $(a,q)$ mesh, with 1000 ions simulated at each grid point, for all three types of defects at $\gamma = +0.02$. The resulting stability maps (Figure~\ref{fig:stability diagram comparison}) reveal a distinct transmission-loss ridge within the stable region for the SRR and SRD cases, with the latter exhibiting the strongest effect. No such localized ridge appears in the DRR configuration; instead, it exhibits a broad reduction in transmission accompanied by diffuse stability boundaries, in contrast to the sharp cutoffs of an ideal quadrupole. Furthermore, all three asymmetric configurations show a substantial contraction of the stable operating region relative to the symmetric QMF.

The distinct transmission characteristics observed in the stability diagrams can be understood by examining how each type of geometric defect modifies the underlying quadrupolar field, adding non-linear contribution from higher-order spatial harmonics. In particular, the emergence of a  resonance dip structures is closely linked to the symmetry-breaking induced changes in the multipole composition of the trapping potential.

\subsection{Nonlinear Resonances}

To identify the field components responsible for the observed transmission anomalies, the simulated electrode potentials were decomposed into their spatial multipole constituents. For the ideal four-fold symmetric geometry ($\eta=1.13$), the expansion is dominated by the quadrupole component ($A_2$), with only the symmetry-allowed higher-order even multipoles ($A_6, A_{10}, ...$) present. The corresponding potential is
\begin{equation}
  \begin{split}
    \Phi_{\gamma=0}(z) = 1.00169z^{2} + 0.00097z^{6} - 0.00247z^{10} - 0.00034z^{14} + \dots
  \end{split}
\end{equation}

The multipole structure changes significantly when the radius perturbation is applied to a diagonally opposite rod pair, reducing the four-fold rotational symmetry of the electrode geometry to two-fold symmetry. The simulated potential decomposition reveals the absence of the odd-order multipoles and the emergence of additional even-order components, most notably the octupole ($A_4$), in addition to the intrinsic multipoles ($A_2$, $A_6$, $A_{10}$, $\ldots$). The corresponding potential configuration is given by
\begin{equation}
  \begin{split}
    \Phi_{\gamma'=+0.02}(z) &= -0.02289 + 0.98035z^{2} + 0.00228z^{4} + 0.00223z^{6} + 0.00051z^{8}\\
    \quad\quad\quad\quad &- 0.00220z^{10} + 0.00008z^{12} - 0.00029z^{14} + \dots
  \end{split}
\end{equation}
The simulated multipole coefficients are in good agreement with the expected symmetry considerations~\cite{sysoev2022balance} and are consistent with the results reported by Dutta \textit{et al}. for the symmetric displacement of diagonally opposite rods.~\cite{Dutta2026}

In contrast, introducing either a single-rod radius variation or a single-rod displacement completely destroys the remaining two-fold rotational symmetry. Consequently, the simulated potential decomposition exhibits the emergence of odd-order multipoles together with the symmetry-allowed even-order components. For identical perturbation magnitudes ($\gamma$), the single-rod radius and displacement defects produce nearly identical multipole spectra. The corresponding potential profiles are expressed as follows:

\begin{equation}
  \begin{split}
    \Phi_{\gamma_d=+0.02}(z) &= -0.01064 - 0.01652z + 0.99143z^{2} - 0.00412z^{3}\\
    \quad\quad\quad\quad &+ 0.00036z^{4} + 0.00067z^{5} + 0.00129z^{6} + 0.00009z^{7}\\
    \quad\quad\quad\quad &+ 0.00023z^{8} + 0.00020z^{9} - 0.00234z^{10} + 0.00006z^{11}\\
    \quad\quad\quad\quad &+ 0.00004z^{12} + 0.00003z^{13} - 0.00031z^{14} + \dots
  \end{split}
\end{equation}
\begin{equation}
  \begin{split}
    \Phi_{\gamma_r=+0.02}(z) &= -0.01152 - 0.01777z + 0.99107z^{2} - 0.00370z^{3}\\
    \quad\quad\quad\quad &+ 0.00118z^{4} + 0.00130z^{5} + 0.00160z^{6} + 0.00019z^{7}\\
    \quad\quad\quad\quad &+ 0.00026z^{8} + 0.00021z^{9} - 0.00233z^{10} + 0.00006z^{11}\\
    \quad\quad\quad\quad &+ 0.00004z^{12} + 0.00003z^{13} - 0.00031z^{14} + \dots
  \end{split}
\end{equation}

\begin{figure}[H]
    \centering
    
    \begin{subfigure}{0.49\textwidth}
        \centering
        \stackinset{l}{-0.2cm}{t}{0.2cm}{\textbf{(a)}}{%
            \includegraphics[width=\textwidth]{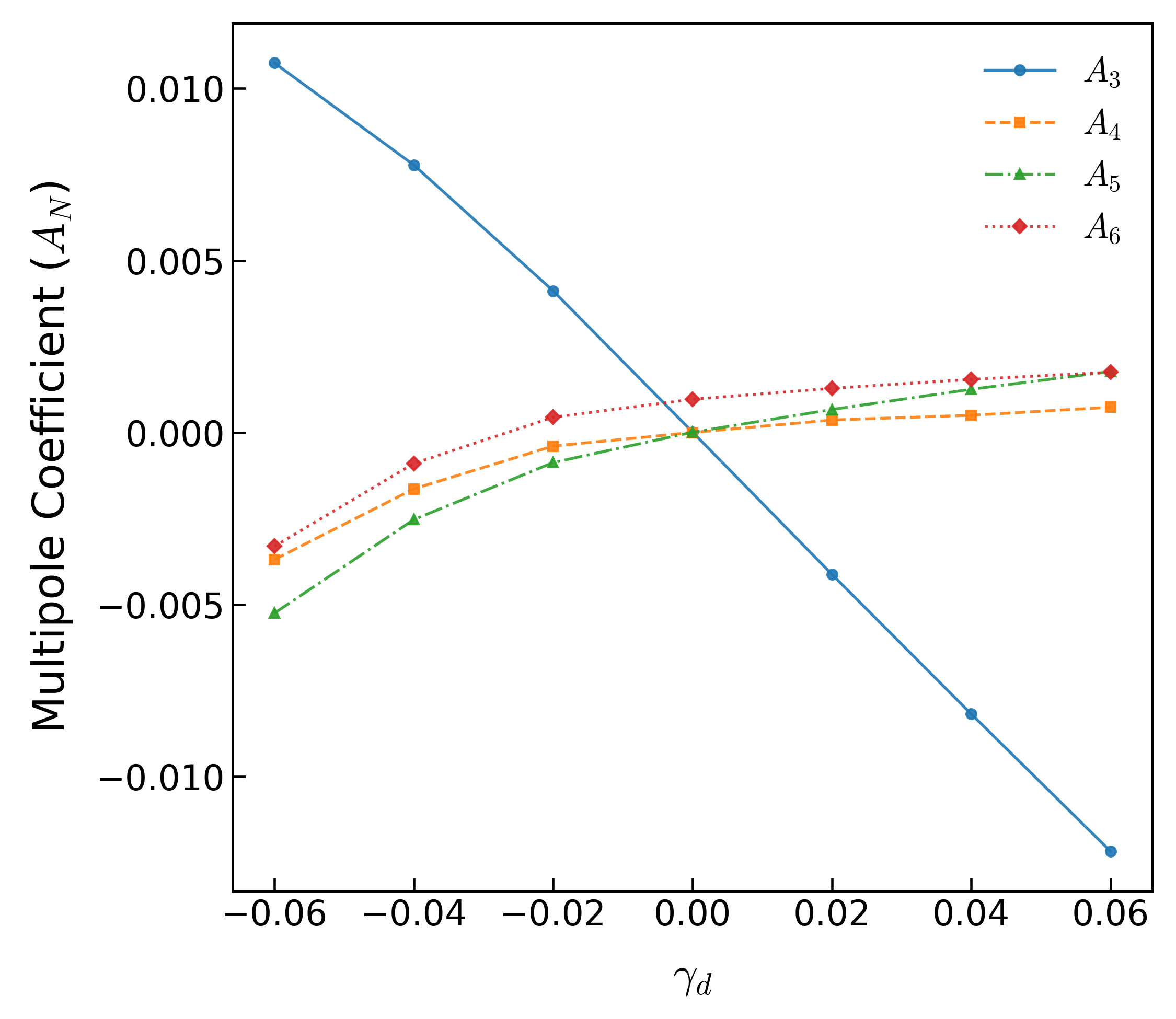}%
        }
        \caption*{} 
        \label{fig:multipole_SRD}
    \end{subfigure}
    \hfill 
    \begin{subfigure}{0.49\textwidth}
        \centering
        \stackinset{l}{-0.2cm}{t}{0.2cm}{\textbf{(b)}}{%
            \includegraphics[width=\textwidth]{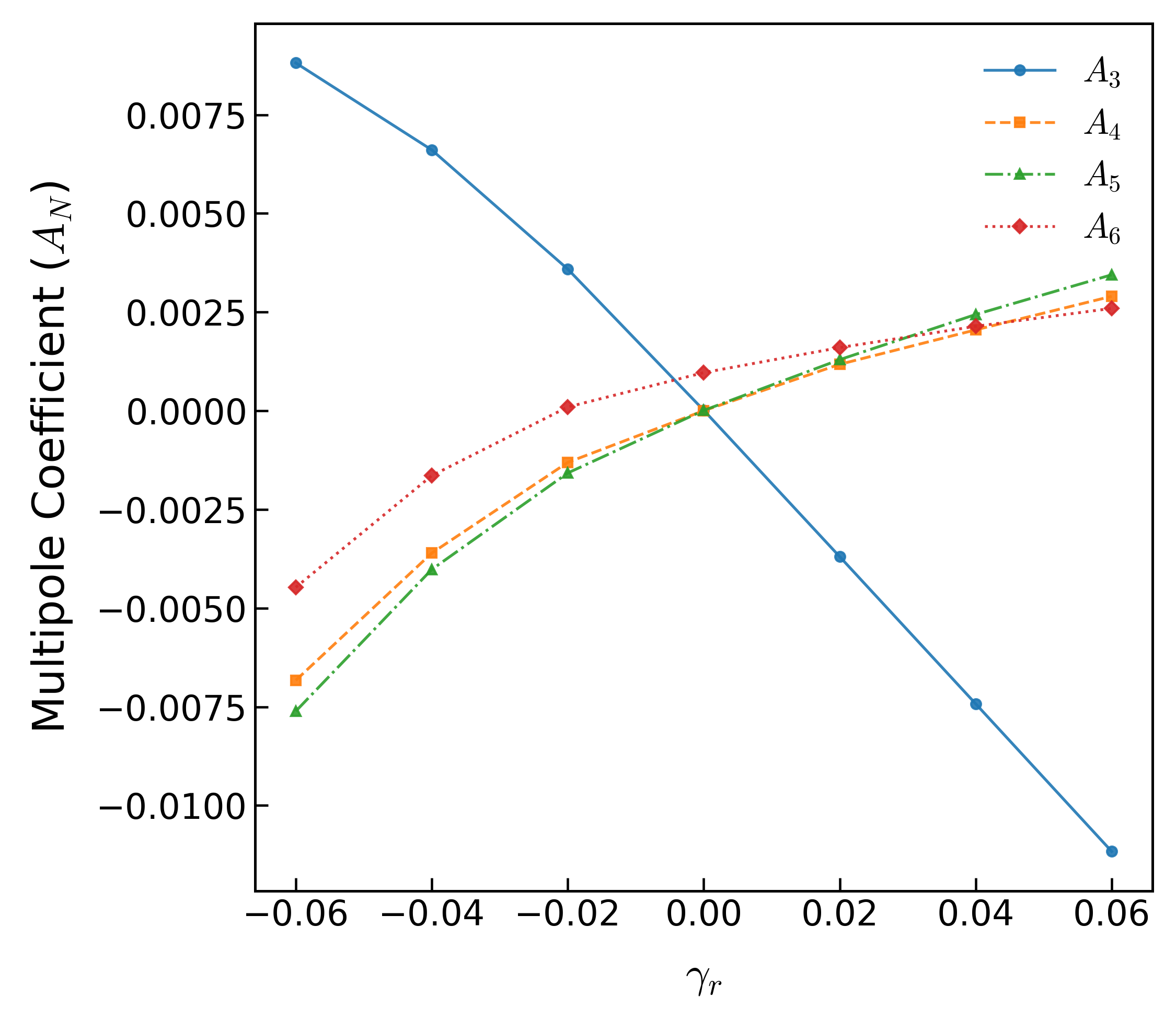}%
        }
        \caption*{} 
        \label{fig:multipole_SRR}
    \end{subfigure}
    
    \vspace{-0.8cm} 

    \caption{Extracted native multipole expansion coefficients for (a) SRD case and (b) SRR case, plotted natively as a function of the geometric asymmetry parameter $\gamma$.}
    \label{fig:multipole_trends_side_by_side}
\end{figure}

As seen in Figure \ref{fig:multipole_trends_side_by_side}, among the multipole coefficients $A_3$, $A_4$, $A_5$, and $A_6$, the $A_3$ mode is clearly dominant in both the SRD and SRR cases. Consequently, it acts as a prominent non-linear distortion field across the entire $\gamma$ range. Most importantly, the $A_3$ term switches sign between positive and negative $\gamma$ values. Simultaneously, the other multipoles show a discrete increase in magnitude with $-\gamma$, but exhibit much less deviation with $+\gamma$.

\begin{figure}[H]
    \centering
    
    \begin{subfigure}{1.0\textwidth}
        \centering
        \stackinset{l}{-0.5cm}{t}{0.2cm}{\textbf{(a)}}{%
            \includegraphics[width=\textwidth]{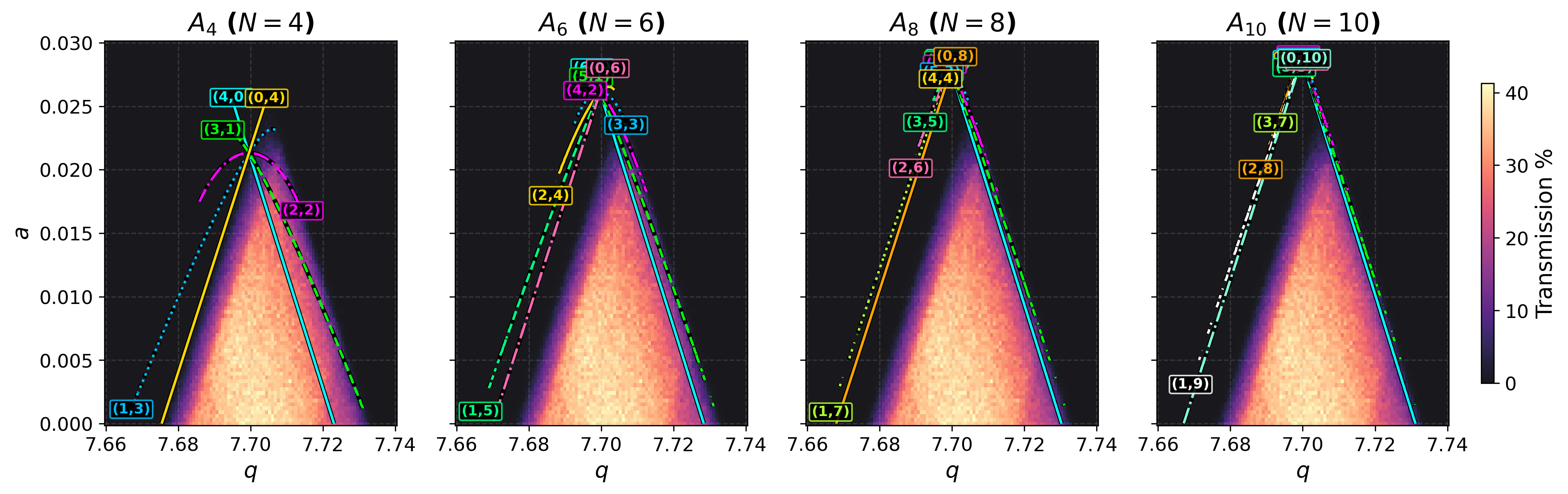}%
        }
        \caption*{} 
        \label{fig:stability_drr_floquet}
    \end{subfigure}
    
    \vspace{-0.8cm} 
    
    \begin{subfigure}{1.0\textwidth}
        \centering
        \stackinset{l}{-0.5cm}{t}{0.2cm}{\textbf{(b)}}{%
            \includegraphics[width=\textwidth]{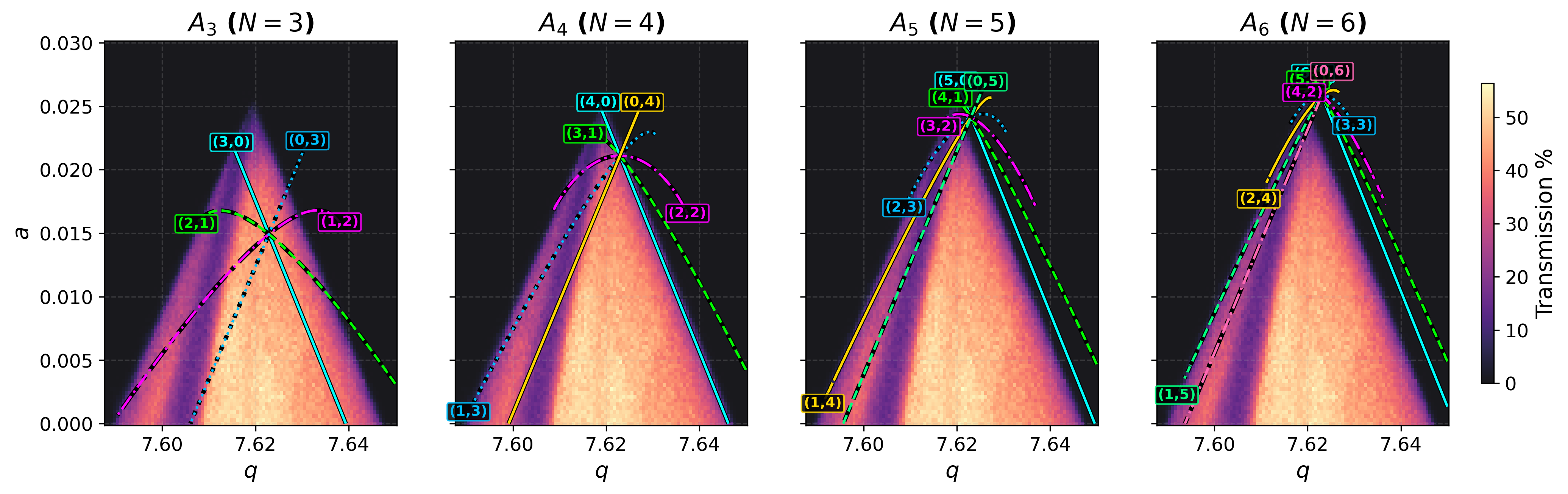}%
        }
        \caption*{} 
        \label{fig:stability_srd_floquet}
    \end{subfigure}
    
    \vspace{-0.8cm}
    
    \begin{subfigure}{1.0\textwidth}
        \centering
        \stackinset{l}{-0.5cm}{t}{0.2cm}{\textbf{(c)}}{%
            \includegraphics[width=\textwidth]{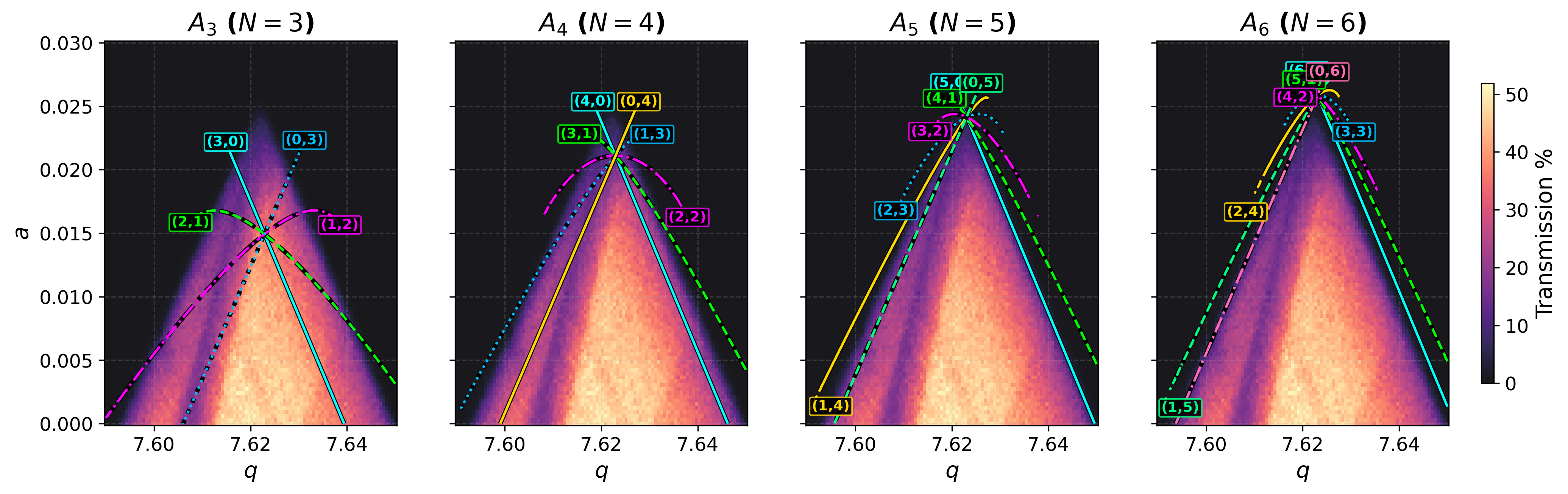}%
        }
        \caption*{} 
        \label{fig:stability_srr_floquet}
    \end{subfigure}

    \vspace{-0.8cm}

    \caption{Overlay of analytical resonance lines on the stability diagram under an asymmetry parameter of $\gamma = 0.02$ computed using the Floquet Transfer Matrix method for (a) DRR ($N = 4, 6, 8, 10$), (b) SRD ($N = 3, 4, 5, 6$), and (c) SRR ($N = 3, 4, 5, 6$) configurations. Designations $(k, l)$ are highlighted at line boundaries.}
    \label{fig:combined_stability_diagrams_floquet_resonance_lines}
\end{figure}

For systematically mapping the manifestation of these structural field components within the highly selective second stability zone, the transmission was recorded over a dense $100 \times 100$ gridpoint array.
Over this transmission background, the secular frequency indices $\beta_x$ and $\beta_y$ were calculated via the effective coordinate scaling framework ($a_{\text{eff}} = a/p, q_{\text{eff}} = q/p$) to accurately mirror the simulated $\gamma = 0.02$ geometric asymmetry profile.\cite{Jana2025RadialAI} Calculated resonance lines are plotted for different kind of asymmetry cases ($\gamma=+0.02$) in Figure~\ref{fig:combined_stability_diagrams_floquet_resonance_lines}.
Within each individual subplot $A_N$, all active resonance line combinations satisfying the sum rule $k + l = N$ are generated and tracked continuously via the continuous second-zone NLR relation:\cite{Du1999}
\begin{equation}
    k (2 - \beta_x) + l (\beta_y - 1) = 1
\end{equation}

These lines are directly superimposed onto the core transmission profile, utilizing distinct colors and linestyles alongside inline coordinate labels to avoid graphical clutter across crowded multi-line bands. The analytical tracking grid reveals a clear structural correlation between the topology of the transmission islands and the high-order resonance boundaries. At lower orders, such as the $A_3$ ($N=3$) and $A_4$ ($N=4$) subplots, the superimposed contours explicitly intersect the primary transmission node. When ion trajectories are scanned along the mass scan lines, cutting these resonance lines, the multi-harmonic driving terms pump dynamic energy directly into the ions' radial motion.\cite{Du1999} This coherent energy transfer triggers rapid growth in the oscillation amplitudes, causing the ions to collide with the physical surfaces of the adjacent rod electrodes before completing their longitudinal flight through the length of the analyzer.

Crucially, a comparative evaluation across the configurations reveals why the driving mechanism fundamentally alters transmission stability. In both the SRD (Figure~\ref{fig:combined_stability_diagrams_floquet_resonance_lines}(b)) and SRR (Figure~\ref{fig:combined_stability_diagrams_floquet_resonance_lines}(c)) cases, the high-order analytical resonance lines demonstrate an intimate geometric alignment with the central ridge of the stability diagram. This structural tracking along the highest-transmission backbone provides a clear explanation for the origin of precursor or anomalous peaks observed in simulations; as the operating scan line sweeps along the apex region, it continuously intersects these closely packed resonance lines, selectively fragmenting the transmission island and trapping narrow, localized high-transmission pockets that manifest as anomalous peaks. The transmission ridges best match with the $A_3$ and $A_4$ NLR lines.

The unique relevance of the DRR configuration shown in Figure~\ref{fig:combined_stability_diagrams_floquet_resonance_lines}(a) becomes evident in this context. Because the symmetric push-pull excitation in the DRR mode inherently suppresses odd-order structural multipoles ($N = 3, 5, \dots$), only the even-order harmonics ($N = 4, 6, 8, 10$) are present to perturb the field. Consequently, the resonance lines do not track directly down the central transmission ridge in the same catastrophic manner seen in the single-rod driven systems. 

As the analysis transitions to these high-order structural harmonics from $A_5$ through $A_{10}$, the density of the $k + l = N$ combinations increases significantly, introducing up to 11 concurrent resonance lines within a single operating block. The visual superimposition demonstrates that while individual high-order lines (e.g., within the $A_8$ or $A_{10}$ spaces) continue to fragment the peripheral boundaries of the operating zone, the core high-transmission backbone remains structurally viable under the dual-rod configuration. As resonance lines coalesce near the apex of the stability diagram, this behavior explains the total transmission loss at the higher $\lambda$ values in Figure \ref{fig:different lambda scan SRR}. This systematic visualization explicitly isolates why manufacturing tolerances impacting rod pairing generate narrow, catastrophic transmission dips at precise coordinates within the second stability zone, providing a direct predictive blueprint for isolating stable operating windows against nonlinear resonance traps.\cite{Du1999}


\section{Conclusion}

This study systematically reveals the distinct electrodynamic mechanisms that separate completely asymmetric manufacturing defects from symmetric paired variations within a linear quadrupole mass filter operated in the second stability zone. By implementing a highly resolved complex spatial multipole expansion framework alongside SIMION ion tracking routines, we established that a structural variance on a singular independent electrode rod—whether introduced via cross-sectional thickness changes ($\gamma_r$) or spatial translations ($\gamma_d$)—provokes a severe and highly localized drop in ion transmission efficiency.
This degradation maps a distinct parametric ridge or valley running through the interior of the second stable zone.\
Conversely, our dual-rod verification control simulations, where a diagonally opposite electrode pair was scaled symmetrically, preserved clean stable bounds across the core of the diagram without any anomalous intensity dips.

This divergence is open to a clear physical explanation rooted in the symmetry of the spatial fields.
While paired anomalies maintain two-fold mirror symmetry and cancel out odd spatial harmonics, a single-rod defect breaks this symmetry completely.
This localized architectural breakdown drives the simultaneous emergence of mixed-parity harmonics, most notably injecting a dominant hexapole ($A_3$) field distortion alongside the symmetric octupole ($A_4$) term.
The determination of secular frequency index confirmed that the presence of this hexapolar field couples the orthogonal focus planes, leading to third-order nonlinear secular resonances.
Ions tracking across these specific coordinates experience exponential energy growth and exit the stable channel via boundary collisions.
These insights provide crucial geometric tolerance guidelines and pole alignment strategies necessary to mitigate resolution caps and preserve peak performance in advanced mass spectrometers operating within higher stability regimes.

\section{Conflict of Interest}

The authors declare no competing financial interest.

\section{Acknowledgment}

N.D. thanks SERB/ANRF India (CRG/2023/001529) and
BRNS India (58/14/21/2023 - BRNS12329) for funding. A.S.
acknowledges DST-Inspire for the research fellowship. N.D. and A.S. thanks IACS for the research infrastructure.


\bibliography{references}

\end{document}